\documentclass[11pt,a4paper]{article}

\usepackage{fullpage}
\usepackage{amsmath}
\usepackage{amssymb}
\usepackage{mathrsfs}
\usepackage{graphicx}
\usepackage{psfrag}
\usepackage[all,poly,curve,knot,cmtip]{xy}
\usepackage{citesort}

\newcommand {\be} {\begin{eqnarray*}}
\newcommand {\ee} {\end{eqnarray*}}
\newcommand {\bea} {\begin{eqnarray}}
\newcommand {\eea} {\end{eqnarray}}

\newcommand{\bm}[1]{\boldsymbol{#1}}

\begin{document}

\title{Supersymmetric isolated horizons in ADS spacetime}
\author{\textbf{Ivan Booth}\footnote{Electronic mail: ibooth at math.mun.ca}\\
\\{\small \it Department of Mathematics and Statistics}\\
{\small \it Memorial University of Newfoundland}\\
{\small \it St. John's, Newfoundland, Canada, A1C 5S7}\\
\\\textbf{Tom$\acute{\mbox{a}}\check{\mbox{s}}$
Liko}\footnote{Electronic mail: liko at gravity.psu.edu}\\
\\{\small \it Department of Physics and Physical Oceanography}\\
{\small \it Memorial University of Newfoundland}\\
{\small \it St. John's, Newfoundland, Canada, A1B 3X7}\\
{\small \it And}\\
{\small \it Institute for Gravitation and the Cosmos}\\
{\small \it Pennsylvania State University}\\
{\small \it University Park, Pennsylvania 16802, USA}}

\maketitle






\begin{abstract}

We discuss various physical aspects of nonextremal, extremal and supersymmetric
black holes in asymptotically anti-de Sitter (ADS) spacetimes.  Specifically,
we discuss how the isolated horizon (IH) framework leads to an ambiguity-free
description of rotating black holes in these spacetimes. We then apply this
framework to investigate the properties of supersymmetric isolated horizons
(SIHs) in four-dimensional  $N=2$ gauged supergravity.  Among other results
we find that they are necessarily extremal, that rotating SIHs must have non-trivial 
electromagnetic fields, and that non-rotating SIHs necessarily have constant curvature 
horizon cross sections and a magnetic (though not electric) charge.

\end{abstract}

\hspace{0.35cm}{\small \textbf{PACS}: 04.50.Gh; 04.70.Bw; 11.30.Pb}


\section{Introduction}

Currently there is a lot of interest in the anti-de Sitter (ADS)/conformal field
theory (CFT) correspondence \cite{maldacena,witten1,witten2,agmoo}.  A significant
amount of effort on the gravity side has been focused on finding charged and
rotating black hole solutions in five-dimensional ADS spacetime, both nonextremal
in general \cite{hht,cejm,hawrea,klesab,clp,cclp} and supersymmetric in particular
\cite{gutrea,klr1,klr2,kunluc1}.

For these black holes, however, there is an ambiguity in how the conserved charges
are defined.  This was first pointed out by Caldarelli \emph{et al} \cite{cck}.
The ambiguity arises because for rotating black holes in ADS spacetime there are
two distinct natural choices for the timelike Killing field.  Defining the charges
with respect to one corresponds to a frame at infinity that is non-rotating and
with the other corresponds to a frame at infinity that is rotating.  The original
motivation for defining the conserved charges using the latter Killing field was
that the corresponding boundary CFT conserved charges satisfy the first law of
thermodynamics \cite{hht}; but this comes at the cost that the bulk conserved
charges do not \cite{cck,gpp1}.  This claim has by now been corrected.  As was
shown in \cite{gpp2}, one can always pass from the bulk conserved charges to the
boundary conserved charges in such a way that both sets seperately satisfy the
first law.  The key to this resolution is that the conserved charges of a rotating
black hole in ADS spacetime have to be measured with respect to the timelike vector
which corresponds to a frame that is non-rotating at infinity.

From the above considerations, it is clear that rotation in ADS spacetime should
be independent of the coordinates that are used.  This is especially crucial when
considering supersymmetric black holes in ADS spacetime (the extremal limit of a
non-rotating ADS black hole results in a naked singularity).  The purpose of this
paper is two-fold: to discuss how the isolated horizon framework provides a
resolution to the above pathology, and to investigate the conditions imposed by
supersymmetry on the corresponding black holes (in four dimensions).

We shall consider the phase space of solutions to the equations of motion for the
Einstein-Maxwell-Chern-Simons (EM-CS) action
\bea
S = \frac{1}{16\pi G_{D}}\int_{\mathcal{M}}\Sigma_{IJ} \wedge \Omega^{IJ}
    - 2\Lambda\bm{\epsilon} - \frac{1}{4}\bm{F} \wedge \star \bm{F}
    - \frac{2\lambda}{3\sqrt{3}}\bm{A} \wedge \bm{F}^{(D-1)/2} \, ,
\label{action1}
\eea
where $\lambda=0$ if $D$ is even and $\lambda=1$ if $D$ is odd.  In this paper,
spacetime indices $a,b,\ldots\in\{0,\ldots D-1\}$ will be raised and lowered using
the metric tensor $g_{ab}$, while internal Lorentz indices
$I,J,\ldots\in\{0,\ldots,D-1\}$ will be raised and lowered using the Minkowski metric
$\eta_{IJ}=\mbox{diag}(-1,1,\ldots,1)$.  The action (\ref{action1}) depends on the
coframe $e^{I}$, the gravitational connection $A_{\phantom{a}J}^{I}$ and the
electromagnetic connection $\bm{A}$.  The coframe determines the metric $g_{ab}$,
$(D-m)$-form $\Sigma_{I_{1}\ldots I_{m}}$ and spacetime volume element
$\epsilon_{a_{1} \ldots a_{D}}$:
\bea
g_{ab} &=& \eta_{IJ}e_{a}^{\phantom{a}I} \otimes e_{b}^{\phantom{a}J}\\
\Sigma_{I_{1}\ldots I_{m}}
       &=& \frac{1}{(D-m)!}\epsilon_{I_{1} \ldots I_{m}I_{m+1} \ldots I_{D}}
           e^{I_{m+1}} \wedge \cdots \wedge e^{I_{D}}\\
\epsilon_{a_{1} \ldots a_{D}}
       &=& \epsilon_{I_{1} \ldots I_{D}}
           e_{a_{1}}^{\phantom{a}I_{1}} \cdots e_{a_{D}}^{\phantom{a}I_{D}} \; .
\eea
Here $\epsilon_{I_{1}\ldots I_{D}}$ is the totally antisymmetric Levi-Civita tensor.
The volume $D$-form $\bm{\epsilon}$ is given by
\bea
\bm{\epsilon} = e^{0} \wedge \cdots \wedge e^{D-1} \; .
\eea
The gravitational connection determines the curvature two-form
\bea
\Omega_{\phantom{a}J}^{I} = dA_{\phantom{a}J}^{I}
+A_{\phantom{a}K}^{I} \wedge A_{\phantom{a}J}^{K}
= \frac{1}{2}R_{\phantom{a}JKL}^{I}e^{K} \wedge e^{L} \, ,
\eea
with $R_{\phantom{a}JKL}^{I}$ as the Riemann tensor.  The electromagnetic connection
$\bm{A}$ determines the curvature
\bea
\bm{F} = d\bm{A} \; .
\eea
The constants appearing in the action (\ref{action1}) are the $D$-dimensional Newton
constant $G_{D}$ and the cosmological constant which in terms of the ADS radius $L$
is given by $\Lambda=-(D-1)(D-2)/(2L^{2})$.

The equations of motion are derived from independently varying the action with respect
to the fields $(e,A,\bm{A})$.  To get the equation of motion for the coframe we note the
identity
\bea
\delta\Sigma_{I_{1} \ldots I_{m}}
= \delta e^{M} \wedge \Sigma_{I_{1} \ldots I_{m}M} \; .
\eea
This leADS to
\bea
\Sigma_{IJK} \wedge \Omega^{JK} + \frac{3}{L^{2}}\Sigma_{I}
= \mathscr{T}_{I} \, ,
\label{eom1}
\eea
where $\mathscr{T}_{I}$ denotes the electromagnetic stress-energy ($D-1$)-form.  The
equation of motion for the connection $A$ is
\bea
\mathscr{D}\Sigma_{IJ} = 0 \, ;
\label{eom2}
\eea
this equation says that the torsion $T^{I}=\mathscr{D}e^{I}$ is zero.  The equation
of motion for the connection $\bm{A}$ is
\bea
d \star \bm{F} - \frac{4(D+1)\lambda}{3\sqrt{3}}\bm{F}^{(D-1)/2} = 0 \; .
\label{eom3}
\eea
The second term in this equation is the contribution due to the CS term in the action.
In even dimensions the equation reduces to the standard Maxwell equation $d \star \bm{F}=0$.

In four dimensions the theory describes the bosonic sector of $N=2$ gauged supergravity,
and in five dimensions the theory describes the bosonic sector of $N=1$ gauged supergravity.
General properties of supersymmetric black holes in ADS supergravity were recently
investigated in \cite{klr2,kunluc1}.  Here we investigate the important issue of rotation
using a more general framework that does not require the spacetime to be globally stationary.

\section{Boundary conditions}

Let us begin with some general remarks concerning the geometrical setup.  We shall
consider a manifold $(\mathcal{M},g_{ab})$ with boundaries.  The conditions that
are imposed on the inner boundary capture the notion of an isolated black hole that
is in local equilibrium with its (possibly) dynamic surroundings.  For details we
refer the reader to \cite{ashkri} and the references therein.  For the boundary
conditions we follow \cite{apv,afk,abl}.

First we give some general comments about the structure of the manifold.  Specifically,
$\mathcal{M}$ is a $D$-dimensional Lorentzian manifold with topology $R\times M$,
contains a $(D-1)$-dimensional null surface $\Delta$ as inner boundary (representing
the horizon), and is bounded by $(D-1)$-dimensional spacelike manifolds $M^{\pm}$ that
extend from $\Delta$ to infinity.  Following \cite{apv}, we assume that the manifold $\mathcal{M}$ can be
conformally completed to an asymptotically ADS spacetime $\widehat{\mathcal{M}}$, where
$\widehat{\mathcal{M}}\cong\mathcal{M}\cup\mathscr{I}$ and $\mathscr{I}$ is a timelike
boundary.  The null surface $\Delta\cong\partial\mathcal{M}$ is by definition the
inner boundary of $\mathcal{M}$.  The topology of $\Delta$ is $R\times\mathbb{S}^{D-2}$
and the topology of $\mathscr{I}$ is $R\times\mathbb{C}^{D-2}$, with $\mathbb{S}^{D-2}$
a compact ($D-2$)-space and $\mathbb{C}^{D-2}$ a ($D-2$)-space.  $M$ is a partial
Cauchy surface such that $\mathbb{S}^{D-2}\cong\Delta\cap M$ and
$\mathbb{C}^{D-2}\cong\mathscr{I}\cap M$.  See Figure 1.
\begin{figure}[t]
\begin{center}
\psfrag{D}{$\Delta$}
\psfrag{I}{$\mathscr{I}$}
\psfrag{Mp}{$M^{+}$}
\psfrag{Mm}{$M^{-}$}
\psfrag{Mi}{$M$}
\psfrag{Sp}{$\phantom{Sp}$}
\psfrag{Sm}{$\phantom{Sm}$}
\psfrag{S}{$\mathbb{S}$}
\psfrag{I}{$\mathscr{I}$}
\psfrag{C}{$\mathbb{C}$}
\psfrag{M}{$\mathcal{M}$}
\includegraphics[width=4.5in]{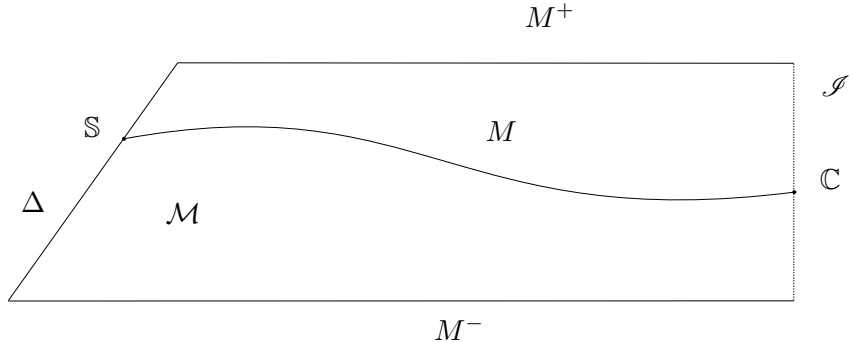}
\caption{The region of the $D$-dimensional spacetime $\mathcal{M}$
being considered has an internal boundary $\Delta$ representing the
event horizon, and is bounded by two $(D-1)$-dimensional spacelike
hypersurfaces $M^{\pm}$ which extend from the inner boundary $\Delta$
to the boundary at infinity $\mathscr{I}$.  $M$ is a partial Cauchy
surface that intersects $\Delta$ in a compact ($D-2$)-space
$\mathbb{S}$ and $\mathscr{I}$ in a ($D-2$)-space $\mathbb{C}$.}
\end{center}
\end{figure}

$\Delta$ is a weakly isolated horizon (WIH).  That is, $\Delta$ is a null surface
and has a degenerate metric $q_{ab}$ with signature $0+\ldots+$ (with $D-2$
nondegenerate spatial directions) along with an equivalence class of null normals $[\ell]$
(defined by $\ell \sim \ell^\prime \Leftrightarrow \ell' = k \ell$ for some constant $k$) such that the
following conditions hold: (a) the expansion $\theta_{(\ell)}$ of $\ell_{a}$
vanishes on $\Delta$; (b) the field equations hold on $\Delta$; (c) the
stress-energy tensor is such that the vector $-T_{\phantom{a}b}^{a}\ell^{b}$ is a
future-directed and causal vector; (d) $\pounds_{\ell}\omega_{a}=0$ and
$\pounds_{\ell}\underleftarrow{\bm{A}}=0$ for all $\ell\in[\ell]$ (see below).

The first three conditions determine the intrinsic geometry of $\Delta$. Since $\ell$ is normal to 
$\Delta$ the associated null congruence is necessarily twist-free and geodesic. By condition
(a) that congruence is non-expanding. Then the Raychaudhuri equation implies that
$T_{ab}\ell^{a}\ell^{b}=-\sigma_{ab}\sigma^{ab}$, with $\sigma_{ab}$ the shear tensor, and 
applying the energy condition (c) we find that $\sigma_{ab} = 0$. Thus, together these conditions
tell us that the intrinsic geometry of $\Delta$ is 
``time-independent'' in the sense that all of its (two-dimensional) cross sections have identical intrinsic geometries. 

Next,  the vanishing of the expansion, twist and shear imply that \cite{afk}
\bea
\nabla_{\!\underleftarrow{a}}\ell_{b}\approx\omega_{a}\ell_{b} \, ,
\label{connectionondelta}
\eea
with ``$\approx$'' denoting equality restricted to $\Delta$ and the underarrow
indicating pull-back to $\Delta$.  Thus the one-form $\omega$ is the natural connection
(in the normal bundle) induced on the horizon.  These conditions also imply that \cite{afk}
\bea
\underleftarrow{\ell \lrcorner \bm{F}} = 0 \; .
\label{pullback1}
\eea
With the field equations (\ref{eom3}) and the Bianchi identity $d\bm{F}=0$, it then
follows that
\bea
\pounds_{\ell}\underleftarrow{\bm{F}}
\approx \ell \lrcorner \underleftarrow{d\bm{F}}
          + d(\underleftarrow{\ell \lrcorner \bm{F}}) = 0 \; .
\label{liepullback}
\eea
This implies that the electric charge is independent of the choice of cross
sections $\mathbb{S}^{D-2}$ \cite{abf}. Similarly (in four-dimensions) the magnetic
charge is also a constant.

From (\ref{connectionondelta}) we find that 
\bea
\ell^{a}\nabla_{\! a}\ell^{b} = (\ell \lrcorner \omega) \ell^{b} \; ,
\eea
and define the surface gravity $\kappa_{(\ell)}=\ell \lrcorner \omega$ as the inaffinity of this geodesic
congruence. Note that it is certainly dependent on specific element of $[ \ell ]$ as under the transformation
$\ell \rightarrow k \ell$:
 \bea
\kappa_{(\ell)} \rightarrow  k \kappa_{(\ell)} \; .
\eea
In addition to the surface gravity, we also define the electromagnetic scalar
potential $\Phi_{(\ell)}=-\ell \lrcorner A$ for each $\ell\in[\ell]$ and this has a similar dependence. 

Now, it turns out that if the first three conditions hold, then one can always find an 
equivalence class $[\ell]$ such that (d) also holds. Hence this last condition does not further
restrict the geometries under discussion, but only the scalings of the null normal. However, 
making such a choice ensures that  \cite{afk}:
\bea
d\kappa_{(\ell)} = d(\ell \lrcorner \omega) = 0
\quad
\mbox{and}
\quad
d\Phi_{(\ell)} = d(\ell \lrcorner \bm{A}) = 0 \; .
\eea
This establishes the zeroth law of WIH mechanics: the surface gravity and scalar
potential are constant on $\Delta$.

\section{Conserved charges: non-rotating reference frame}

The derivation of the conserved charges involves first finding the symplectic
structure on the covariant phase space $\bm{\Gamma}$ consisting of solutions
$(e,A,\bm{A})$ to the field equations (\ref{eom1}), (\ref{eom2}) and (\ref{eom3})
on $\mathcal{M}$.  Once we have a suitable (closed and conserved) symplectic
two-form
\bea
\bm{\Omega} \equiv \bm{\Omega}(\delta_{1},\delta_{2}) \, ,
\eea
the conserved charges are obtained by evolving the system with respect to appropriate
vector fields (symmetries).  Two sets of conserved charges arise this way: those at
$\mathscr{I}$ corresponding to a non-rotating frame at infinity and those at $\Delta$
corresponding to the horizon charges that satisfy the first law of black-hole mechanics.

The antisymmetrized second variation of the surface term gives the symplectic current.
Integrating this current over a spacelike hypersurface $M$ gives the bulk symplectic
structure (with the choice of $M$ being arbitrary).  This two-form, however, is generally
not conserved.  This is due to the fact that the symplectic current can ``leak'' across
the horizon.  In order to obtain a symplectic structure that is conserved on $\bm{\Gamma}$
we need to find the pull-back of the current to $\Delta$ and add the integral of this
term to the symplectic structure.

To find the conserved charges for any system, one examines the canonical transformations
that are generated by the corresponding Hamiltonians.  For some smooth vector field $\xi$
that preserves the boundary conditions of Section 2 and any vector field $\delta$ that is
tangent to $\bm{\Gamma}$, it follows that the necessary and sufficient condition for
$\delta_{\xi}$ to be a phase space symmetry (i.e. that $\pounds_{\delta_\xi}\bm{\Omega}=0$
on $\bm{\Gamma}$) is that
\bea
\bm{\Omega}(\delta,\delta_{\xi}) = \delta\mathscr{H}_{\xi} \, ,
\eea
where $\mathscr{H}_{\xi}$ is the Hamiltonian generating the infinitesimal diffeomorphism
and is given by
\bea
\mathscr{H}_{\xi} = \mathscr{Q}_{\xi}^{(\mathscr{I})} - \mathscr{Q}_{\xi}^{(\Delta)} \; .
\label{hamiltonian}
\eea
The conserved charges for WIHs in asymptotically ADS spacetimes with no matter fields
were derived in \cite{apv}.  Inclusion of matter fields does not involve any significant
modifications to the conserved charges.

As was shown in Appendix B of \cite{him}, inclusion of antisymmetric tensor fields in the
action does not contribute anything to the charges at $\mathscr{I}$ because the fields
fall off too quickly.  Therefore the charges at infinity for EM-CS theory are precisely
the ones that were derived in \cite{apv}; these are the Ashtekar-Magnon-Das (AMD) charges
\cite{ashmag,ashdas}:
\bea
\mathscr{Q}_{\xi}^{(\mathscr{I})} = \frac{L}{8\pi G_{D}}\oint_{\mathbb{C}^{D-2}}
                                    \widetilde{E}_{ab}\xi^{a}\tilde{u}^{b}\bm{\tilde{\varepsilon}} \, ,
\label{amdcharges}
\eea
with $\tilde{u}^{a}$ the unit timelike normal to $\mathbb{C}^{D-2}$, $\bm{\tilde{\varepsilon}}$
the area form on $\mathbb{C}^{D-2}$ and $\widetilde{E}_{ab}$ the leading-order electric part of
the Weyl tensor.  Explicitly we have that
\bea
\widetilde{E}_{ab} = \frac{1}{D-3}\Omega^{3-D}\widetilde{C}_{abcd}\tilde{n}^{c}\tilde{n}^{d} \, ,
\eea
where $\tilde{n}^{a}=\tilde{\nabla}^{a}\Omega$, and $\Omega$ is the conformal factor defined
via $\tilde{g}_{ab}=\Omega^{2}g_{ab}$ which relates the unphysical metric $\tilde{g}_{ab}$ on
$\widehat{\mathcal{M}}$ and the physical metric $g_{ab}$ on $\mathcal{M}$.

Gibbons \emph{et al} \cite{gpp1} showed that the asymptotic time translation Killing field
for an exact solution has to be chosen in such a way that the frame at infinity is non-rotating.
If this is done then the AMD charge evaluated for the solution will result in an expression for
mass that satisfies the first law.  Moreover, Gibbons \emph{et al} \cite{gpp2} showed that
using this definition for the asymptotic time translation has to be used for a consistent
transition to the conserved charges of the boundary CFT.

At the horizon, inclusion of Maxwell fields gives rise to an electric charge \cite{likboo}
\bea
\mathcal{Q} = \frac{1}{8\pi G_{D}}\oint_{\mathbb{S}^{D-2}}\bm{\Phi} \; .
\label{charge}
\eea
where $\bm{\Phi}$ is the electromagnetic charge density:
\bea
\bm{\Phi} = \star\bm{F} - \frac{4(D+1)\lambda}{3\sqrt{3}}\bm{A} \wedge \bm{F}^{(D-3)/2} \, ,
\label{chargedensity}
\eea
(not to be confused with the Coulomb potential $\Phi$).  Due to the presence of the CS term
(when $\lambda=1$), the charge $\mathcal{Q}$ may fail to be gauge invariant if the horizon
has a non-trivial topology.  Other conserved quantities at the horizon include the horizon
entropy and angular momenta \cite{likboo}
\bea
\mathcal{S} &=& \frac{1}{4G_{D}}\oint_{\mathbb{S}^{D-2}}\bm{\tilde{\epsilon}}
\label{entropy}\\
\mathcal{J}_{\iota} &=& \frac{1}{8\pi G_{D}}\oint_{\mathbb{S}^{D-2}}
                    \left[(\phi_{\iota} \lrcorner \omega)\bm{\tilde{\epsilon}}
                    + (\phi_{\iota} \lrcorner \bm{A})\bm{\Phi}\right] \, ,
\label{angularmomentum}
\eea
where $\phi_{\iota}$ are rotational Killing fields, and we define the
area element of the cross section $\mathbb{S}^{D-2}$ of the horizon
\bea
\bm{\tilde{\epsilon}} = \vartheta^{(1)} \wedge \cdots \wedge \vartheta^{(D-2)} \; .
\eea
The index $\iota\in\{1,\ldots,\lfloor (D-1)/2 \rfloor\}$ is a rotation index and
corresponds to $\lfloor (D-1)/2 \rfloor$ independent rotation parameters in $D$
dimensions and are given by the Casimirs of the rotation group $SO(D-1)$, where
$\lfloor \cdot \rfloor$ denotes ``integer value of''.

It was shown in \cite{likboo} that these charges satisfy the first law.  The angular
momenta contain contributions from gravitational as well as electromagnetic fields,
referred to here as $\mathcal{J}_{\rm Grav}$ and $\mathcal{J}_{\rm EM}$ respectively.
If $\phi$ is the restriction to $\Delta$ of a \emph{global} rotational Killing field
$\varphi$ contained in $\mathcal{M}$, then the electromagnetic contribution to
(\ref{angularmomentum}) can be interpreted as angular momentum of the electromagnetic
radiation in the bulk \cite{abl}.  This is because the bulk integral can be written
as the sum of two surface terms -- one at $\Delta$ and one at $\mathscr{I}$; the
latter surface term is zero due to the fall off conditions at $\mathscr{I}$.
Therefore the condition for a WIH to be non-rotating is that $\mathcal{J}_{\rm Grav}=0$
for all rotational Killing fields. This will be discussed in greater detail in the next
section.

\section{Supersymmetric isolated horizons}

Until now we have discussed the mechanics of  WIHs in arbitrary dimensions.  We now specialize
to supersymmetric horizons in ADS spacetime and in particular we focus on the bosonic sector
of four-dimensional $N=2$ gauged supergravity.  In this case, black holes are solutions to the
bosonic equations of motion and so the fermion fields vanish. By definition, supersymmetric
solutions are invariant under the full supersymmetry transformations.  This means that for
black hole solutions, these transformations should leave the fermion fields unchanged (and
vanishing). Therefore any such black hole solutions must admit a Killing spinor field. 

For full stationary black hole solutions such as those discussed in \cite{kosper,calkle1}, the
Killing spinor gives rise to a (timelike) time-translation Killing vector field in the region
outside of the black hole horizon. However, in the quasilocal spirit of the isolated horizon
programme we will only assume the existence of a Killing spinor \emph{on the horizon itself}.
In this case the spinor will generate a null geodesic vector field that has vanishing twist,
shear, and expansion and this is an allowed $\ell$ on the WIH.

In order to proceed we now restrict our attention to fully isolated horizons (IHs).  These are
WIHs for which there is a scaling of the null normals for which the commutator
$[\pounds_{\ell},\mathcal{D}]=0$, where $\mathcal{D}$ is the intrinsic covariant derivative on
the horizon.  In contrast to condition (d) for WIHs, this condition cannot always be met and geometrically
such horizons not only have time-invariant intrinsic geometry, they also have time-invariant extrinsic geometry.
That said it is clear that this condition similarly fixes $\ell$ only up to a constant scaling. As such it does not 
uniquely determine the value of the surface gravity $\kappa_{(\ell)}$ but does fix its sign. In particular this allows 
us to invariantly say whether or not $\kappa_{(\ell)}$ vanishes. This then gives rise to an invariant characterization
of extremality that is intrinsic to the horizon: an \emph{extremal isolated horizon} is an IH on
which $\kappa_{(\ell)} = 0$.  Further discussion of this notion of extremality and how it relates
to other characterizations may be found in \cite{boofai1}. 

Finally we define a \emph{supersymmetric isolated horizon} (SIH) as an IH on which the null
vector generated by the Killing spinor coincides (up to a free constant) with the preferred null
vector field arising from the IH structure. As we shall now see these are necessarily extremal 
as well as having restricted geometry, rotation, and matter fields.

For four-dimensional $N=2$ gauged supergravity, we shall employ the conventions of
\cite{calkle2}.  The corresponding (bosonic) action is
\bea
S = \frac{1}{16\pi G_{4}}\int_{\mathcal{M}}\Sigma_{IJ} \wedge \Omega^{IJ}
    + \frac{6}{L^{2}}\bm{\epsilon} - \frac{1}{4}\bm{F} \wedge \star \bm{F} \; .
\label{em}
\eea
The necessary and sufficient condition for supersymmetry with vanishing fermion fields
is that there exists a Killing spinor $\epsilon^{\alpha}$ such that
\bea
\left[\nabla_{\! a} + \frac{i}{4}F_{bc}\gamma^{bc}\gamma_{a}
      +\frac{1}{L}\gamma_{a}\right]\epsilon = 0 \; .
\label{cond1}
\eea
Here, $\gamma^{a}$ are a set of gamma matrices that satisfy the anticommutation rule
\bea
\gamma^{a}\gamma^{b} + \gamma^{b}\gamma^{a} = 2\eta^{ab}
\eea
and the antisymmetry product
\bea
\gamma_{abcd} = \epsilon_{abcd} \; .
\eea
$\gamma_{a_{1}\ldots a_{D}}$ denotes the antisymmetrized product of $D$ gamma matrices.
The spinor $\epsilon$ satisfies the reality condition
\bea
\bar{\epsilon} = i(\epsilon)^{\dagger}\gamma_{0} \, ;
\eea
overbar denotes complex conjugation and $\dagger$ denotes Hermitian conjugation.

From $\epsilon$ one can construct five bosonic bilinears $f$, $g$, $V^{a}$, $W^{a}$ and
$\Psi^{ab}=\Psi^{[ab]}$ where
\bea
f = \bar{\epsilon}\epsilon \, ,
\quad
g = i\bar{\epsilon}\gamma^{5}\epsilon \, ,
\quad
V^{a} = \bar{\epsilon}\gamma^{a}\epsilon \, ,
\quad
W^{a} = i\bar{\epsilon}\gamma^{5}\gamma^{a}\epsilon \, ,
\quad
\Psi^{ab} = \bar{\epsilon}\gamma^{ab}\epsilon \; .
\eea
These are inter-related by several algebraic relations (from the Fierz identities) and differential
equations (from the Killing equation (\ref{cond1})) \cite{calkle2}.  For our purposes the significant
ones are:
\bea
V_{a}V^{a}
&=& -W_{a}W^{a} = -(f^{2} + g^{2}) \, , \label{tnvector1} \\
V^a W_a &=& 0 \label{VdW} \, , \\
g W_{a}
&=& \Psi_{ab} V^b \label{PsiV} \, , \\
f \Psi_{ab} &=& -  \epsilon_{abcd} V^c W^d + \frac{1}{2} g \epsilon_{abcd} \Psi^{cd} \, ,\label{VwW} \\
\nabla_a f
& = & F_{ab} V^b \label{Fl} \, , \\
\nabla_a g
& = & - \frac{1}{L} W_a - \frac{1}{2} \epsilon_{abcd} V^b F^{cd} \label{Wdef} \, ,\\
\nabla_a V_b
&=& \frac{1}{L} \Psi_{ab}
    - f F_{ab} + \frac{g}{2} \epsilon_{abcd} F^{cd} \label{Psi1} \, ,\\
 \nabla_a A_b &=& - \frac{g}{L} g_{a b} - F_{(a}^{\; \; c} \epsilon_{b)cde} \Psi^{de}
  + \frac{1}{4} g_{ab} \epsilon_{cdef} 
  F^{cd} \Psi^{ef}  \, \mbox{and} \label{gradA}
 \\   
\nabla_c \Psi_{ab}
&=& \frac{2}{L} g_{c [a} V_{b]} + 2 F_{[a}^{ \; \; d}\epsilon_{b]dce}W^{e}
    + F_{c}^{\; \; d}\epsilon_{dabe}W^{e} + g_{c[a}\epsilon_{b]def}W^{d}F^{ef} \label{OmegaRel} \; . 
\eea
These are general relations for the existence of a Killing spinor in spacetime.  Although the Killing spinor
may exist in a neighbourhood of the horizon, we only require that it exist \emph{on the horizon itself}.
Henceforth we specialize by setting $f=g=0$ and at the same time require that the relations hold on $\Delta$.
Thus, the differential equations (\ref{Fl})-(\ref{OmegaRel}) are only required to hold when the derivatives
are pulled-back onto the horizon. 

With $f=g=0$,  equation (\ref{tnvector1}) implies that $V^a$ and $W^a$ are both null. On an 
SIH we identify $\ell^a = V^a$ and so condition (\ref{connectionondelta}) together with the differential
constraint (\ref{Psi1}) implies that
\bea
\nabla_{\underleftarrow{a}} \ell_{b} =  \omega_a \ell_b = \frac{1}{L}\Psi_{\underleftarrow{a}b}  \, ,
\label{psipullback}
\eea
and using the skew-symmetry of $\Psi_{ab}$ we can write
\bea
\Psi_{ab} = L (\omega_a \ell_b - \omega_b \ell_a) \, . \label{Psi2}
\eea
Then by equation (\ref{PsiV}) 
 \bea
\ell \lrcorner \omega = 0 \Leftrightarrow \kappa_{(\ell)}=0 \, .\label{kap0}
\eea
Thus, an SIH is necessarily extremal. 

For ease of presentation we now assume that the SIH is foliated into spacelike two-surfaces 
$\Delta_{v}$. One can always construct such a foliation (and its labelling) so that the associated null 
normal $n\equiv dv$ satisfies $\ell \lrcorner n = -1$ \cite{abl2}. Then the two-metric and 
area form on the $\Delta_{v}$ can be written as
\bea
\tilde{q}_{ab} = g_{ab} + \ell_{a}n_{b} + \ell_{b}n_{a}
\quad
\mbox{and}
\quad
\tilde{\epsilon}_{cd} = - \ell^{a}n^{b}\epsilon_{abcd}  
\eea
respectively.  Now we note that $\omega$ can be written as
\bea
\omega_{a} = -\kappa_{(\ell)}n_{a} + \tilde{\omega}_{a} \, ,
\eea
with $\tilde{\omega}$ the pull-back to $\Delta_{v}$ of $\omega$.  Then with $\kappa_{(\ell)} = 0$
it follows that $\omega_{a}=\tilde{\omega}_{a}$ and hence $\omega_{a} \in T^\star (\Delta_v)$.  Finally, with
respect to this foliation, the usual restriction (\ref{pullback1}) and (redundantly) equation (\ref{Fl}) implies
that the electromagnetic field takes the form
\bea
F_{ab} = E_\perp (\ell_a n_b - n_a \ell_b) +  B_\perp \tilde{\epsilon}_{ab} + 
(\tilde{X}_a \ell_b - \tilde{X}_b \ell_a) \, , \label{F}
\eea
on $\Delta$. Here, $E_\perp$ and $B_\perp$ are the electric and magnetic fluxes through the surface and 
$\tilde{X}^a \in T(\Delta_{v})$ describes flows of electromagnetic radiation along (but not through) the horizon.

With these preliminaries in hand we can consider the properties of SIHs in asymptotically ADS spacetimes.
In particular, it was previously shown \cite{likboo} that in the absence of a cosmological constant, SIHs
are necessarily non-rotating with $\omega = 0$ and so it is natural to consider how the addition of a
negative cosmological constant affects the rotation properties.  First, relations (\ref{VdW}) and
(\ref{VwW}) tell us that
\bea
W^a = L \beta V^a
\eea
for some function $\beta$ (the factor of $L$ has been included for later convenience). Then
the pull-back of (\ref{Wdef}) trivially vanishes without giving us any new information but 
(\ref{gradA}) provides a differential equation for $\beta$ on each $\Delta_v$
\bea
d_a \beta + \beta \tilde{\omega}_a = B_\perp \tilde{\omega}_a - E_\perp \tilde{\epsilon}_a^{\; \; b}
 \tilde{\omega}_b \, , 
\eea
where $d_a$ is the intrinsic covariant derivative on $\Delta_v$,
along with its time-invariance: $\pounds_\ell \beta = 0$. 

Next applying the various properties of extremal IHs, one can show that the pull-back of (\ref{OmegaRel})
is 
\bea
\nabla_{\! \underleftarrow{c}} \Psi_{a b} = 2 L \left( \frac{1}{L^2} - \beta B_\perp \right) 
\tilde{q}_{c[a} \ell_{b]} + 2 L \beta E_\perp \tilde{\epsilon}_{c[a} \ell_{b]} \, ,  
\eea
and combining this with (\ref{Psi2}) we find that 
\bea
d_a \tilde{\omega}_b + \tilde{\omega}_a \tilde{\omega}_b
=  \left( \frac{1}{L^2} - \beta B_\perp \right) \tilde{q}_{ab} + \beta E_\perp \tilde{\epsilon}_{ab} \; .
\label{dtom}
\eea

Now as was seen in (\ref{angularmomentum}), the gravitational angular momentum associated with a rotational
Killing field $\phi^a$ is 
\bea
\mathcal{J}_{\rm Grav} = \frac{1}{8 \pi G_4}\oint_{\Delta_v} \bm{\tilde{\epsilon}} \phi \lrcorner \tilde{\omega} \, ,
\eea
and so a necessary condition for non-zero angular momentum is a non-vanishing {rotation one-form} $\tilde{\omega}_a$.
That said, this is not quite sufficient as it is possible for a non-vanishing $\phi \lrcorner \tilde{\omega}$ to
integrate to zero.  For example, consider the case where $\Delta_{v}$ has topology $S^2$ and $\phi^a$ is a Killing
field (and so divergence-free). Then for some function $\zeta$  we can write $\phi^a = \tilde{\epsilon}^{ab} d_b \zeta$
and
\bea
\oint_{\Delta_{v}} \bm{\tilde{\epsilon}} \phi \lrcorner \tilde{\omega} = \oint_{\Delta_{v}}  \zeta d \tilde{\omega} \; .
\eea
Thus, for all closed rotational one-forms ($d \tilde{\omega} = 0$) the associated gravitational angular momentum
will vanish.  As such, it is standard in the isolated horizon literature (see e.g. \cite{abl}) to take
$d\tilde{\omega}\neq0$ as the defining characteristic of a rotating isolated horizon. In our case
\bea
 d_{[a} \tilde{\omega}_{b]} =  \beta E_\perp \tilde{\epsilon}_{ab} \; , 
\eea
and so an SIH is rotating if and only if $\beta E_\perp \neq 0$. Thus, a rotating horizon must have a
non-trivial electromagnetic field. This is in agreement with known exact solutions: rotating supersymmetric
Kerr-Newmann-AdS black holes as well as those with cylindrical or higher genus horizons all have non-trivial
EM fields \cite{calkle2}.

\section{Summary and discussion}

Let us summarize the role that IHs play in ADS spacetime, and the resulting conclusions
that we can draw from them regarding the generic properties of nonextremal, extremal, and
supersymmetric black holes. 

The IH framework provides a coherent physical picture whereby two sets of conserved charges
arise in ADS spacetime: the charges measured at infinity and the local charges measured at
the horizon.  The local conserved charges at the horizon then satisfy the first law.  When
evaluated on exact solutions to the field equations, the charges at infinity correspond to
asymptotic symmetries that are measured with respect to a \emph{non-rotating} frame at infinity.

As was the case for spacetimes with no cosmological constant, supersymmetric isolated horizons
in ADS spacetime have vanishing surface gravity and so are always extremal.  However, in contrast
to the asymptotically flat case, we found that ADS SIHs in four dimensions can be either rotating
or non-rotating with strong constraints linking the rotation to the electromagnetic and Killing spinor
fields.  The use of these constraints to classify SIHs in ADS spacetime will appear in a future work;
here we will give a taste of their application by considering the case when $\tilde{\omega}=0$.  Then,
the Maxwell equations along with the extremal IH conditions tell us that $E_\perp$ and $B_\perp$ are
both constant in time ($\pounds_\ell E_\perp = \pounds_\ell B_\perp = 0$) and
\bea
d_a B_\perp + \tilde{\epsilon}_a^{\; \; b} d_b E_\perp = 0 \, . 
\eea
Hence $E_\perp$ and $B_\perp$ are also constant on each $\Delta_v$. Next the supersymmetry constraint
(\ref{dtom}) says that
\bea
\beta B_\perp = \frac{1}{L^2} \; \; \mbox{and} \; \; \beta E_\perp = 0  \, .  
\eea
Thus, $E_\perp = 0$ while $B_\perp \neq 0$ -- that is, these SIHs necessarily have magnetic, but not
electric, charges.  Further, applying the extremality condition from \cite{likboo,boofai1}:
\bea
\frac{1}{2} \mathcal{R} &=& d_a \tilde{\omega}^a + \tilde{\omega}_a \tilde{\omega}^a + T_{ab} \ell^a n^b
                            - \frac{3}{L^2} \label{extremality}\\
&=& B^2_\perp - \frac{3}{L^2} \; .
\eea
It is clear that the two-dimensional Ricci  curvature $\mathcal{R}$ of the $\Delta_v$ is constant in this
case -- unfortunately the sign of that curvature does not seem to be determined by the equations.  Consulting
a listing of exact supersymmetric black hole solutions \cite{calkle2} we see that such solutions are known:
specifically there is a supersymmetric asymptotically ADS black hole in four dimensions which can be non-rotating
if the horizon cross sections have genus $g>1$. As prescribed by our formalism, these solutions have magnetic but
not electric charge. 

The extremality condition (\ref{extremality}) also gives us information in the case of rotating SIHs.
Specifically, integrating and applying the Gauss-Bonnet theorem we find that:
\bea
\mathcal{A}_{\Delta_v}
= \frac{L^2}{{3}}\left[4\pi(g-1) + \oint_{\Delta_v}   \mspace{-10mu} \bm{\tilde{\epsilon}}
(E_\perp^2+B_\perp^2 + \|\tilde{\omega}\|^{2})\right] \; ,
\label{inequality}
\eea
where $\mathcal{A}_{\Delta_v}\equiv\oint_{\Delta_{v}}\bm{\tilde{\epsilon}}$ is the area of the horizon
cross sections. More generally for non-extremal horizons the equality above becomes a ``$\geq$" and so this
equality becomes a bound. Then the maximum allowed angular momentum is bound by the genus and area of the
horizon; see \cite{boofai1,MaxRot} for discussions of the corresponding result for asymptotically flat
spacetimes and appendix B of \cite{boofai1} for a particular discussion of Kerr-ADS. 

Alternatively, reversing the inequality, one can view it as bounding the allowed area of isolated horizons
from below by the scale of the cosmological curvature and the genus of the horizon: higher genus horizons
necessarily have larger areas. Similar bounds have previously been discovered for stationary ADS black holes
\cite{gibbons,woolgar,caigal}.

The description of ADS black holes presented here is somewhat different from the description of black holes in globally
stationary spacetimes where an ambiguity appears that manifests itself as a choice of whether the conserved charges are
measured with respect to a frame at infinity that is rotating or non-rotating.  This ambiguity does not appear in the IH
framework essentially because the conserved charges of the black hole are measured \emph{at the horizon}, and the
corresponding first law is intrinsic to the horizon with no mixture of quantities there and at infinity!

{\bf Note added.}  After this work was completed, it was brought to our attention that equation (\ref{extremality})
has been solved recently in \cite{kunluc2} for vacuum gravity in the context of near-horizon geometries.

\section*{Acknowledgements}

We thank Dumitru Astefanesei for comments during the initial stages of this work.  The authors were supported by the
Natural Sciences and Engineering Research Council of Canada.

\end{document}